\documentclass{ws-procs9x6}

\begin{document}

\title{Using the X-ray Emission Lines of Seyfert 2 AGN to Measure Abundance
Ratios }

\author{MARIO A. JIMENEZ-GARATE and TOAN KHU\footnote{\uppercase{T}his work
was supported by \uppercase{NASA} contract \uppercase{NAS} 8-01129.}}

\address{MIT Center for Space Research, 70 Vassar St., NE80-6009 \\ 
Cambridge, MA, 02139, USA. E-mail: mario@space.mit.edu}


\maketitle

\abstracts{We measure the metal abundance ratios in the X-ray photoionized 
gas located near the narrow line region of a sample of Seyfert 2 AGN.
The high-resolution X-ray spectra 
observed with the Chandra high- and low-energy transmission grating 
spectrometers are compared with models of
the resonant scattering and recombination
emission from a plasma in thermal balance, 
and with multiple temperature zones.  
The abundance ratios in the sample are close to the Solar values, 
with slight over-abundances of N in NGC 1068, and of Ne in 
NGC~4151.  Our X-ray spectral models use fewer degrees of freedom 
than previous works. }

{\bf Motivation}. Our goal is to use X-ray abundance measurements to cross-calibrate
with the optical spectra of quasars.  These
optical spectra are used to measure the star formation history of
the early Universe.\cite{hamann}

\begin{figure}[ht]
\centerline{\epsfxsize=4.7in\epsfbox{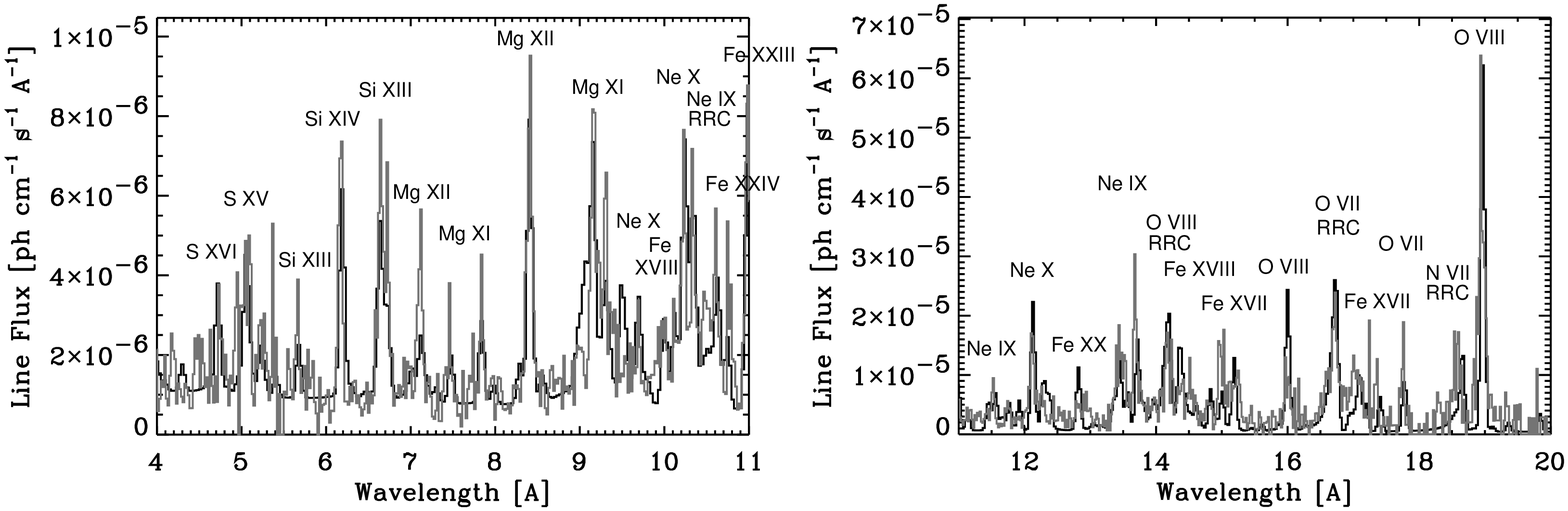}}
\caption{A portion of the X-ray spectrum of NGC~1068 measured with
the {\it Chandra} high-energy transmission grating (grey),
and model fit (black). \label{fig}}
\end{figure}

{\bf To Determine the Abundance Ratios,}
we fit the Seyfert 2 spectra with models of photoionized plasmas, and
then search for deviations in the data from Solar abundances,
as shown in Fig. \ref{fig}.  The fluxes of
radiative recombination continuum (RRC) features and the radiative
recombination (RR) forbidden lines depend little on radiation transfer
effects, so they are the most reliable abundance indicators.
At low continuum optical depths, the recombination emission 
flux scales linearly with abundances.  Our model
allows us to fit the RRC and RR fluxes by accounting for
the broad ionization distribution in the plasma. 

{\bf X-ray Emission Line Model.}
We model a photoionized plasma in ionization equilibrium and thermal
balance with a grid of zones, each with 
ionization parameter $\log \xi = 1.0,1.5,...4.5$ (in c.g.s. units).  
The XSTAR plasma code yields the charge state
distribution of each zone.\cite{kallman}  
In order to calculate the recombination emission,
we use the photoelectric cross sections from Ref.~\refcite{vernery},
and the recombination rates calculated with the HULLAC atomic code,\cite{klap} 
provided by Ref. \refcite{liedahl}.  To calculate the
resonance line fluxes, we use the
oscillator strengths from Ref. \refcite{verner}.
We assume the ionizing spectrum of a radio-quiet quasar,\cite{elvis} 
with an X-ray power-law and high-energy exponential cutoff.
We assume a power-law distribution of column density as
a function of $\xi$, so $N_H ( \xi  ) \propto (\log_{10} \xi)^\beta$,
with $\beta$ a free parameter.\cite{jimen} 
The advantages of this model are that 1) it provides a simple fit to the
ionization distribution, 2) it has only five free parameters,
instead of the several dozen parameters in other models,\cite{kink}
and 3) it uses accurate atomic data.  

{\bf Conclusion}.  The abundances in
NGC~1068, NGC~4051, and NGC~4507 do not deviate much from 
the Solar values. Nitrogen is over-abundant by a factor of $\lesssim 2$
in NGC~1068, and neon is over-abundant in NGC~4051. 
There are no signatures of starburst activity.
Further work is needed to quantify the systematic and statistical errors,
and to compare with optical spectra.

\end{document}